\documentclass[journal]{IEEEtran}
\usepackage{cite}

\ifCLASSINFOpdf
   \usepackage[pdftex]{graphicx}
\else
   \usepackage[dvips]{graphicx}
\fi
\usepackage[cmex10]{amsmath}
\usepackage{amssymb}
\usepackage{algorithmic}

\usepackage[bookmarks=false,dvipdfm,colorlinks,,linkcolor=blue]{hyperref}
\usepackage[thmmarks]{ntheorem}
{
\newtheorem{thm}{Theorem} 
\newtheorem{defn}{Definition} 
\newtheorem{lem}{Lemma} 
}

\usepackage{newfloat}
\usepackage{algorithmic}
\usepackage{threeparttable}

\newcounter{AlgorithmFlag}
\newcommand{\AlgorithmFlag}{\textbf{Algorithm}\refstepcounter{AlgorithmFlag} {\bf \arabic{AlgorithmFlag}}~}

\DeclareFloatingEnvironment[
    fileext=loa,
    listname=List of Algorithms,
    name=ALGORITHM,
    placement=tbhp,
]{algorithm}

\newcounter{algleo}
\newlength{\lefttab}
\newlength{\numberoffset}
\newenvironment{algleo}%
  {\trivlist
   \topsep=0pt\parsep=0pt\itemsep=0pt
   \def\li{\item\refstepcounter{algleo}\makebox[0.8em][r]{\thealgleo\hspace{\numberoffset}}
       \hangafter1\hangindent1.8em\noindent}%
   \def\linonumber{\item\makebox[0.8em][r]{\hspace{\numberoffset}}
       \hangafter1\hangindent1.8em\noindent}%
   \addtolength{\lefttab}{1.25em}
   \addtolength{\numberoffset}{1.25em}
   \leftskip=\lefttab}%
  {\endtrivlist}

\begin{document}
%
\title{Efficient Bit Sifting Scheme of Post-processing in Quantum Key Distribution}
%
%
%

\author{Qiong~Li,
        Dan~Le,
        Xianyan~Wu,
        Xiamu~Niu$^*$,
        and~Hong~Guo
\thanks{Q. Li, D. Le, X. Wu and X. Niu are with the School
of Computer Science and Technology, Harbin Institute of Technology, Harbin, 150001 China.}
\thanks{H.Guo is with the School of Electronics Engineering and Computer Science and Center for Quantum Information Technology, Peking University, Beijing, 100871, China.}
\thanks{X. Niu and D. Le is with the e-mail: xm.niu@hit.edu.cn and ledan@hit.edu.cn, respectively.}}

%
%

\markboth{}%
{}
%



\maketitle

\begin{abstract}
Bit sifting is an important step in the post-processing of Quantum Key Distribution (QKD) whose function is to sift out the undetected original keys. The communication traffic of bit sifting has essential impact on the net secure key rate of a practical QKD system, and it is facing unprecedented challenges with the fast increase of the repetition frequency of quantum channel. In this paper, we present an efficient bit sifting scheme whose core is a lossless source coding algorithm. Both theoretical analysis and experimental results demonstrate that the performance of our scheme is approaching the Shannon limit. Our scheme can greatly decrease the communication traffic of the post-processing of a QKD system, which means it can decrease the secure key consumption for classical channel authentication and increase the net secure key rate of the QKD system. Meanwhile, it can relieve the storage pressure of the system greatly, especially the device at Alice side. Some recommendations on the application of our scheme to some representative practical QKD systems are also provided.
\end{abstract}

\begin{IEEEkeywords}
Quantum cryptography, post-processing, bit sifting, source coding, unconditionally secure authentication.
\end{IEEEkeywords}

%
\IEEEpeerreviewmaketitle

\section{\label{sec:introduction}Introduction}
%
%
%
%
\IEEEPARstart{T}{he} quantum key distribution (QKD) is the most developed branch of quantum cryptography, whose security is based on the principles of quantum mechanics. It can not only enhance the security of traditional symmetric/asymmetric cryptographic systems, but also construct an information-theoretic secure cryptographic system by combing with Vernam one-time pad cipher\cite{shannon1948bell}. QKD comprises two phases: the transmission of the photons over the quantum channel and the post-processing over the authenticated classical channel. In the first phase, by transmitting the modulated photons, Alice and Bob obtain a partially shared bit-string, so called original key. A representative high performance QKD system can transmit original keys at rates in the order of Gbps. In the second phase, by performing sifting, error reconciliation and privacy amplification in an authenticated classical channel, Alice and Bob obtain the identical and unconditionally secure key, so called secure key. The highest secure key rate is about 1Mbps according to the published literatures\cite{tanaka2012high, dixon2010continuous}. The essential procedures of post-processing include sifting, error reconciliation and private amplification. Every procedure of post-processing is responsible for the dramatic loss of the key rate between the original key and secure key. The function of the first procedure is to sift out the undetected original keys and the original keys whose preparation and measurement basis are incompatible, which is also named bit sifting and basis sifting respectively. The loss due to basis sifting depends on the protocol gain of the QKD system. For example, the protocol gain for BB84 protocol\cite{bennett1984quantum} is 0.5, which is introduced by Bennett and Brassard in 1984 and still the most widely used QKD protocol at present. The loss due to bit sifting is determined by the count rate of the QKD system. The count rate is also called detection probability in some publications. The loss caused by the private amplification is the cost to pay for decreasing Eve's knowledge about the secure key to almost zero. Most of Eve's knowledge is obtained from the exchanged messages during the error reconciliation. So far, most studies on post-processing focus on improving the secure key rate via increasing the reconciliation efficiency, which means decreasing the amount of interactive information during the error reconciliation. While the huge amount of interactive messages  during sifting has not drawn enough attentions.

The reason that we should study the method to decrease the communication traffic of sifting is mainly related to the key consumption due to the authentication of classical channel. One of the basic assumptions of the security analysis for QKD protocols is that there is an authenticated classical channel between Alice and Bob\cite{mayers1996quantum, mayers2001unconditional, ben2005universal}. However, the classical channel in a QKD system cannot be authenticated by itself unless we authenticate all interactive messages between Alice and Bob by employing an unconditionally secure authentication algorithm, i.e. the family of almost strongly universal hash functions based algorithm at the cost of some key consumption. For the first round of the QKD system, a pre-shared key must be available, which is exchanged through a secret channel, such as face to face or other ways. For the following rounds, a part of the secure key generated by the QKD system is used as the authentication key. In order to maximize the net secure key rate after the withdrawal by the authentication in a practical QKD system, it is essential to minimize the key consumption of authentication. The consumed key lengths of some representative authentication algorithms are listed in Table \ref{tbl-The consumed key length of some authentication schemes}, as functions of the security parameter and the authenticated message length. It can be found that the consumed key length monotonically increases with the message length $m$. Therefore, we must try to reduce the communication traffic as much as possible. In the post-processing of QKD, the sifting procedure needs to communicate much more than the other procedures. Lin etc. presented a software implementation of post-processing of QKD, and declared that sifting procedure needed the most network resource in 2009\cite{lin2009implementation}. So it is of great meaning to study how to decrease the communication traffic of sifting.
\begin{table}[htbp]
\centering
\caption{The consumed key lengths of some representative authentication algorithms for given security parameter $\varepsilon$ and message length $m$. } \label{tbl-The consumed key length of some authentication schemes}
\begin{tabular}{cc}
 \hline
 { Authentication algorithm} & { Consumed Key Length}\\
 \hline
 den Boer \cite{den1993simple} &  $\approx  - 2{\log _2}\varepsilon  + 2{\log _2}m$\\
 Bierbrauer etc. \cite{bierbrauer1994families}& $\approx  - 3{\log _2}\varepsilon  + 2{\log _2}m$\\
 Krawczyk \cite{krawczyk1994lfsr}& $- 3{\log _2}\varepsilon  + 3{\log _2}\left( {1 + 2m} \right) + 1$\\
 Abidin etc. \cite{abidin2012new}& $ - 4{\log _2}\varepsilon  + 3{\log _2}m + 8$\\
 \hline
\end{tabular}
\end{table}

With the fast increase of the repetition frequency of quantum channel, post-processing devices are facing unprecedented challenges. Taking BB84 protocol as an example, the input data rate of sifting procedure at Alice side is twice the repetition frequency and three times at Bob side. For decoy BB84 protocol\cite{hwang2003quantum}, the input data rate of sifting is even more than that of BB84. To date, the repetition frequency of a high speed QKD system has been up to about ten GHz\cite{tanaka2012high}, so the post-processing devices are facing huge storage pressure. In 2007, Mink indicated that the storage of sifting was one bottleneck for his QKD system\cite{mink2007custom}, whose repetition frequency is 3.125Gbps. It is noted that although the input data rate at Bob side is more than that at Alice side, Bob can immediately sift out the undetected original keys whose amount is far more than the amount of the detected, while Alice cannot sift out them until Bob announces which original keys he has detected, so the storage pressure of the device at Alice side is much heavier. Therefore, the method to decrease the communication traffic of sifting should not only have good compression performance but also be performed as fast as possible so that Alice could remove the undetected original keys from her buffer in time.

Although the post-processing has drawn much attention since the mid-1990s, only very few researchers study the sifting procedure. In 2010, in order to save communication traffic, Kollmitzer etc. stated that Bob could inform Alice the detection position represented by $\left\lceil {{{\log }_2}m} \right\rceil$ bits, where $m$ is the number of original keys to be processed\cite{kollmitzer2010applied}. The scheme can reduce the amount of exchanged messages to some extent, but the compression efficiency is far from the optimum. This scheme was implemented by Li etc. in 2012\cite{li2012design}. In 2014, Walenta etc. declared that the sifting should be performed as fast as possible to allow Alice to sift out the undetected and incompatible original keys to avoid buffer overflow. They also pointed out that the amount of bits exchanged during sifting should be kept as small as possible due to the authentication cost \cite{walenta2014fast}. Their sifting scheme is to encode the detection time indexes between two adjacent detection events at Bob side. Their compression efficiency is less than twice the Shannon limit when the count rate is between $10^{-4}$ and $10^{-1}$, while the performance falls sharply when the count rate is out of the limit.

The key point to decrease the communication traffic of sifting procedure is to reduce the amount of interactive messages during bit sifting step to a maximum extent owing to the following two reasons. As mentioned above, the sifting consists of bit sifting and basis sifting. The function of bit sifting is to get rid of the undetected original keys and the basis sifting aims to sift out the original keys whose bases are incompatible. On one hand, the amount of interactive messages of bit sifting is far more than that of basis sifting. On the other hand, from the point of information theory, the great redundancy due to the very low count rate makes it possible to decrease the amount of interactive messages of bit sifting significantly. While the interactive messages during basis sifting can hardly be compressed because of the low redundancy due to the completely random basis selection at both Alice and Bob sides. Hence, we only focus our study on bit sifting in this paper.

In this paper, we firstly present a lossless source coding based bit sifting scheme. Considering the expected codelength of the source coding algorithm as the optimization object, an efficient iteration algorithm is proposed to solve the optimization problem. Both theoretical analysis and experimental results demonstrate that the performance of our scheme is approaching the Shannon limit and also better than the competitive scheme within the entire reasonable interval of count rate. Besides, some suggestions on how to apply our scheme to some representative practical QKD systems are provided.

The rest of this paper is organized as follows.  In section \ref{sec:Description-Sift}, some preliminaries are presented. In section \ref{sec:A coding scheme for bit sifting step}, the proposed lossless source coding based bit sifting scheme and the theoretical analysis on its performance are discussed in detail. The experimental results and analysis are presented in section \ref{sec:Experimental-Results}. Finally, some conclusions are drawn in section \ref{sec:Conclusion}.

\section{\label{sec:Description-Sift}preliminaries}
In the section, some theoretical bases for proposed bit sifting scheme are introduced briefly.

\subsection{convergence for series}
\begin{defn}
(\textbf{Absolutely convergent})\cite{johnsonbaugh2012foundations}. Given a series $\sum {a_k}$, we may form a new series $\sum {|a_k|}$. If the new series is
convergent, then we say that the original series $\sum {a_k}$ is absolutely convergent.
\end{defn}
\begin{thm}   \cite{johnsonbaugh2012foundations}
Suppose that $a_k \ge 0$ for all $k \ge 1$. Then the series $\sum {a_k}$ either converges or diverges to $+\infty$. Especially, if it converges, then it converges absolutely.
\label{THM_Convergence for series}
\end{thm}
\begin{defn}
(\textbf{Rearrangement of series})\cite{johnsonbaugh2012foundations}. Suppose that $\sum_{k=1}^{+\infty}{a_k}$ is a given series. Let $\{n_k\}$ be a sequence of positive integers such that each positive integer occurs exactly once in the sequence. That is, there exists a bijective map $f: \mathbb{N}^+ \to \mathbb{N}^+ $ with $f(k)=n_k, k \in \mathbb{N}^+$, so that each term in the series $\sum_{k=1}^{+\infty}{b_k}(b_k=a_{n_k})$ is also a term in $\sum_{k=1}^{+\infty}{a_k}$, but occurs in different order. The series $\sum_{k=1}^{+\infty}{b_k}$ is called a rearrangement of $\sum_{k=1}^{+\infty}{a_k}$.
\end{defn}
\begin{thm}
\cite{johnsonbaugh2012foundations}. If $\sum_{k=1}^{+\infty}{a_k}$ converges absolutely with sum $s$, then every series $\sum_{k=1}^{+\infty}{b_k}$ obtained by rearranging its terms also converges absolutely to the same sum $s$.
\label{THM_Rearrangement of absolutely convergent series}
\end{thm}

\subsection{Shannon's limit of source coding}
\begin{defn}(\textbf{Entropy})\cite{cover2006elements}. Let $X$ be a discrete random variable with alphabet $\mathbb{X}$ and probability mass function $p\left( x \right),x \in \mathbb{X}$. the entropy $H(X)$ of the discrete random variable $X$ is defined by
$H(X) =  - \sum\limits_{x \in \mathbb{X}} {p(x)\log_b p(x)}$. In this paper, $b$ is set to 2.
\label{defn:Entropy}
\end{defn}

\begin{defn}
(\textbf{Source code})\cite{cover2006elements}. A source code $C$ for a random variable $X$ is a mapping from $\mathbb{X}$, the range of $X$, to $D^*$, the set of finite length strings of symbols from a $D\text{-ary}$ alphabet. Let $C\left(x\right)$ denote the codeword corresponding to $x$ and $l(x)$ denote the length of $C(x )$.
\label{defn:Instantaneous Code}
\end{defn}

\begin{defn}
(\textbf{Expected Codelength})\cite{cover2006elements}. The expected codelength (also called average codelength) $\overline L$ of a source code $C$ for a random variable $X$ with probability mass function $p\left(x\right)$ is given by ${\overline L} = \sum\limits_{x \in \mathbb{X} } {p\left( x \right)l\left( x \right)}.$
\label{defn:Compression Ratio}
\end{defn}

\begin{thm}
\cite{cover2006elements}. Given a discrete memoryless source of entropy $H(X)$, the average codelength $\overline L$ for any distortionless source encoding scheme is bounded by $\overline L  \ge H\left( X \right)$.
\label{THM_Cover_Elements}
\end{thm}

According to Theorem  \ref{THM_Cover_Elements}, $H(X)$ is the theoretical lower bound of the average codelength per source letter, so the definition of compression efficiency is defined as follows.

\begin{defn}
(\textbf{Compression Efficiency}) Suppose $C$ is a lossless source code of the discrete random variable $X$, and $\overline L$ is the expected codelength of $C$. The compression efficiency of the source code $C$ is given by $f=\frac{\overline L}{H(X)}.$
\label{defn:Compression Effective}
\end{defn}

In this paper the indicator of compression efficiency is used to evaluate the compression performance of a source coding algorithm. According to the Theorem \ref{THM_Cover_Elements}, the more $f$ is closer to 1, the better the source code. Since the entropy $H(X)$ is a constant for a given information source represented by the random variable $X$, the smaller expected codelength $\overline L$ indicates the better compression efficiency.

\section{\label{sec:A coding scheme for bit sifting step}proposed bit sifting scheme}
\subsection{description of bit sifting scheme}
\begin{figure*}[!t]
\centering
\includegraphics{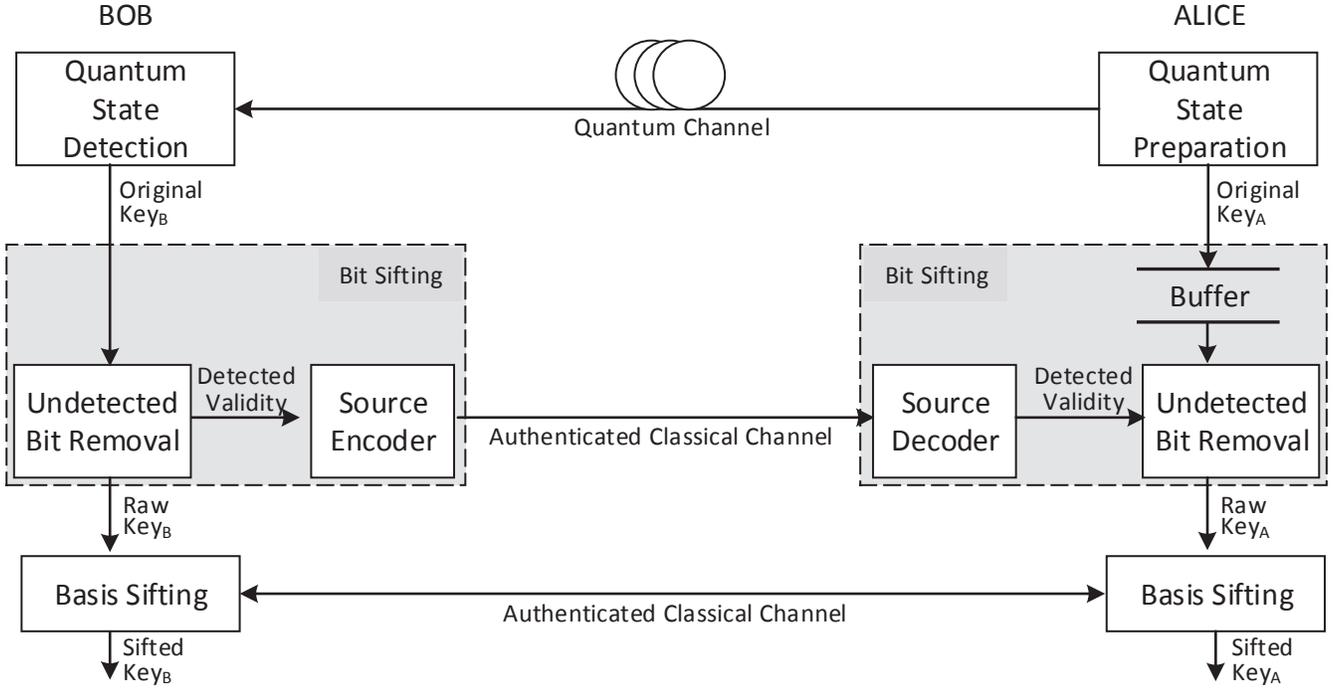}
\caption{\label{fig:bit_sifting_schematic} The bit sifting schematic diagram with the preceding and following steps.}
\end{figure*}

The schematic diagram of the proposed bit sifting scheme  with the preceding and following steps is shown in Fig.\ref{fig:bit_sifting_schematic}. A basic QKD protocol starts with the preparation, transmission and detection of  a random sequence  modulated photons, also called qubits or quantum states, which are transferred into the original key at both sides. The original key constitutes the input of QKD post-processing system. Since a large fraction of qubits cannot be detected due to the loss of the transmission and the imperfection of the detection device, Bob needs to announce the detected validity of each original key.  As mentioned above, the data amount of the announcements is extremely large, which requires a huge secure key consumption for the corresponding authentication. In order to save the secure key consumption, a source encoder and decoder is designed in our bit sifting scheme at Bob and Alice side respectively. The optimal compression performance of the source coding algorithm is pursued to minimize the secure key consumption for the authentication of bit sifting. Since Alice has to buffer all original keys until she receives the validity announcement from Bob, the storage pressure would be too much to bear if the source encoding and decoding cannot be implemented in real time. Therefore,  another desired performance of the source coding algorithm is low computation complexity.

\subsection{description of the MZRL source coding algorithm}
Generally, the announcement is a binary string, in which the value of each bit indicates the detected validity of the corresponding original key. Without loss of generality, we assume that "0" represents the case of undetected, and "1" represents the case of detected. Since the number of photons in one pulse, the noise of quantum channel, and the response of detection device are all almost random, the detected validity is nearly random. So the announcement of detected validities can be considered as a binary memoryless information source which is just the object that we need to compress via a source coding.

Considering that the number of "0" in the binary string is far more than the number of "1", a modified zero run length (MZRL) source coding algorithm is designed. First of all let us recall the traditional zero run length coding. Suppose that there is a binary string "0010001100000001", the traditional zero run length coding result would be "2-3-0-7". Such simple coding algorithm is not completely suitable in the context of QKD. Since a QKD system may run continuously, the length of zero run may be any element from the set of natural number, i.e. $\left\{ {0,1,2, \cdots ,+\infty} \right\}$. That is to say that the binary information source is transferred to a non-binary source with infinite and countable source letters. While it is not realistic to represent infinite numbers in a practical system. In many cases, the run lengths larger than a preset threshold are truncated because the probabilities of the big run lengths are usually so small that they can be neglected in some error tolerant applications. But such truncation scheme does not fit for QKD since "lossless" is the basic requirement for the bit sifting scheme and any error is not acceptable.

\begin{figure}[htbp]
\includegraphics[width=0.5\textwidth]{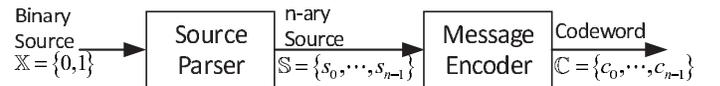}
\caption{\label{fig:MZRL coding scheme} Encoding schematic diagram of MZRL codes.}
\end{figure}

To losslessly represent infinite possible run lengths by using finite resources, we design the MZRL algorithm based on a straightforward and efficient idea, i.e. segmentation. The encoding schematic diagram of MZRL algorithm is shown in the Fig.\ref{fig:MZRL coding scheme}.

The function of the \emph{Source Parser} in Fig.\ref{fig:MZRL coding scheme} is to divide the binary source output sequence into messages, which are the objects to be assigned codewords by the \emph{Message Encoder}. In Fig.\ref{fig:MZRL coding scheme}, the output of \emph{Source Parser} are $n$ variable-length messages. The function of the \emph{Message Encoder} is to map each message into a codeword. To simplify the decoding operation, the length of every codeword is set to $\left\lceil {{{\log }_2}n} \right\rceil $. The output messages of the \emph{Source Parser} and their corresponding codewords produced by the \emph{Message Encoder} in the MZRL coding are presented in Table \ref{tbl-MZRL codes}.
\begin{table}[htbp]
\caption{MZRL codes.} \label{tbl-MZRL codes}
\centering
\begin{tabular}{cc}
\hline
 {Source Message} & {Codeword}\\
 \hline
 ${s_0} = {\rm{"1"}}$ & ${c_0} = 00 \cdots 0$\\
 ${s_1} = "01"$ & ${c_1} = 00 \cdots 1$\\
 $\vdots$ & $\vdots$\\
 ${s_{n - 2}} = "\underbrace {00 \cdots 0}_{n - 2{\text{ }}0'\text{s}}1"$ & ${c_{n - 2}}$\\
 ${s_{n - 1}} = "{\underbrace {00 \cdots 00}_{n - {1\text{ }}0'\text{s}}}"$ & ${c_{n - 1}}$\\
 \hline
\end{tabular}
\end{table}

As shown in Table \ref{tbl-MZRL codes}, the definition of messages are the same as the traditional zero run length coding except the $n{\rm{th}}$ message ${s_{n - 1}}$. The first $n-1$ messages $s_i$ follow the same pattern whose length is $i+1$ and the last digit is 1. But the $n{\rm{th}}$  message $s_{n-1}$ is a sequence of all 0's of length $n-1$. It is obvious that $n$ must be greater than or equal to 2.

It is clear that the codeword $c_0$ to $c_{n-2}$ can represent the run length from 0 to $n-2$. How to represent a run length that is greater than or equal to $n-1$ is the next problem we need to solve. Our method is to segment the long binary sequence into one or more $s_{n-1}$'s and one message $s_i$ ($0 \le i \le n - 2$), which can be represented by $c_{n-1}$ and $c_i$ ($0 \le i \le n - 2$) respectively. That is to say, for an arbitrary zero run length $RL(0) = m * (n - 1) + i$, where $m,i \in \mathbb{N}{\text{ and }} 0 \le i \le n - 2$, the codeword sequence is ${c_{n - 1}}_{_0}{c_{n - 1}}_{_1} \cdots {c_{n - 1}}_{_{m - 1}}{c_i}$. For example, if $n = 4$, the MZRL codeword sequence for the binary string "0010001100000001" is "${c_2}{c_3}{c_0}{c_0}{c_3}{c_3}{c_1}$".

According to Table \ref{tbl-MZRL codes}, both encoding and decoding are quite simple and efficient. For the encoder, the \emph{Source Parser}  stores letters from the \emph{Binary Source} until it sees that these letters form a valid message as defined in Table \ref{tbl-MZRL codes} and the \emph{Message Encoder} outputs the corresponding codeword. For the decoder, it decodes the received codeword $c_i$ to the corresponding message $s_i$ which is just the final output of the decoder. Since MZRL is a non-singular fix-length code, it is by nature an instantaneous code, which means the end of a codeword is immediately recognizable and a codeword can be decoded without reference to future codewords. Such property makes the decoding of MZRL more efficient.

The simple encoding and decoding principles of MZRL guarantee that the algorithm can be implemented very fast. Except the computation complexity, what we care about most in the scenario of QKD is the compression efficiency of the source coding algorithm. Since the shorter expected codelength indicates the better compression efficiency for a given information source, we explore the optimal expected codelength of MZRL algorithm in the following sections.

\subsection{expected codelength of MZRL codes}
Suppose that the count rate of a QKD system is $q$, which means the probability of "1" and "0" are $q$ and $1-q$ for the binary source $X$ in Fig.\ref{fig:MZRL coding scheme}. It is easy to deduct that the probability of the zero run length $l$ is
\begin{equation}
P\left(l\right) = {\left(1-q\right)^{l}}q.
\end{equation}

For any given $i \in \left\{ {0,1, \cdots ,n - 2} \right\}$, the message $s_i$ only appears once when the zero run lengths $l \in \left\{ {l|l=m(n-1)+i, m \in \mathbb{N}} \right\}$. Therefore $P\left(s_i\right)$ is given by
\begin{eqnarray}
{P}\left( s_i \right)&=&\mathop \sum \limits_{m = 0}^{ + \infty } {P}\left( {m(n-1)+i} \right)\nonumber\\
&=& \frac{{{{\left( {1 - q} \right)}^{i + 1}}q}}{{1 - {{\left( {1 - q} \right)}^n} - q}},\;\;\forall i \in \left\{ {0,1,...,n - 2} \right\}.
\label{eq:ps-i}
\end{eqnarray}
While the message $s_{n-1}$ would appears $\lfloor { \frac{l}{n-1} } \rfloor$ times when the zero run lengths $l \in \left\{ {l\left| {l \ge n-1\; \cap \;l \in {\mathbb{N} }} \right.} \right\}$. So $P\left(s_{n-1}\right)$ is given by
\begin{align}
{P}\left( s_{n-1} \right) = \sum\limits_{l = n-1}^{ + \infty } {\left\lfloor {\frac{l}{n-1}} \right\rfloor {P}\left( l \right)}.
\label{eq:Psn-1-org}
\end{align}
Since ${P}\left( {{l}} \right) \ge 0$ and
\begin{eqnarray}
{P}\left( {s_{n-1}} \right) &=& \sum\limits_{l = n-1}^{ + \infty } {\left\lfloor {\frac{l}{n-1}} \right\rfloor {P}\left( l \right)} \nonumber\\
 &\le& \sum\limits_{l = 0}^{ + \infty } {l{P}\left( l \right)} \nonumber\\
 &=& \frac{1}{q} - 1,
\end{eqnarray}
the series ${P}\left( {s_{n-1}} \right)$ in the Eq.(\ref{eq:Psn-1-org}) converges absolutely according to the Theorem \ref{THM_Convergence for series}. According to the Theorem \ref{THM_Rearrangement of absolutely convergent series}, any rearranged series of the series ${P}\left( {s_{n-1}} \right)$ also converges absolutely to the same sum. In order to compute the sum of ${P}\left( {s_{n-1}} \right)$, it is rearranged as follows,
\begin{eqnarray}
{P}\left( {s_{n-1}} \right) &=& \sum\limits_{m = 1}^{ + \infty } {m\sum\limits_{k = 0}^{n - 2} {P\left( {m\left( {n - 1} \right) + k} \right)} }\nonumber\\
&=& \frac{{{{(1 - q)}^n}}}{{1 - {{(1 - q)}^n} - q}}.
\label{eq:ps-n-1}
\end{eqnarray}

Since the codelength of the codeword $c_i$ corresponding to the message $s_i$ is a constant $\left\lceil {{{\log }_2}n} \right\rceil$, and the probability mass function $P\left(s_i\right)$ is given by Eq.(\ref{eq:ps-i}) and Eq.(\ref{eq:ps-n-1}), the expected codelength of MZRL code $C$ for the n-ary source, i.e. the random variable $S$, is given by
\begin{eqnarray}
{\overline{L}_C} &=& \sum\limits_{i = 0}^{n-1} {P\left( {{s_i}} \right){\left\lceil {{{\log }_2}n} \right\rceil}} \nonumber\\
 &=& \frac{{\left\lceil {{{\log }_2}n} \right\rceil }}{{1 - {{(1 - q)}^{n - 1}}}}
\label{eq:ls-aver}
\end{eqnarray}
according to the Definition \ref{defn:Compression Ratio}. To obtain the expected codelength for the binary source, i.e. the random variable $X$, we also need to compute the average length of the source message of $S$ by Eq.(\ref{eq:L-D})
\begin{eqnarray}
{{\overline{L}_S}} &=& \sum\limits_{i = 0}^{n - 2} {P({s_i})\left( {i + 1} \right) + P({s_{n - 1}})\left( {n - 1} \right)}\nonumber\\
&=& \frac{1}{{q}}.
\label{eq:L-D}
\end{eqnarray}
Hence the expected codelength for the binary source $X$ is
\begin{eqnarray}
\overline{L}\left(n\right) &=& \frac{{ {{\overline{L}_C}}}}{{ {{\overline{L}_S}}}}\nonumber\\
&=& \frac{{q\left\lceil {{{\log }_2}n} \right\rceil }}{{1 - {{(1 - q)}^{n - 1}}}}.
\label{eq:R-fixed-length}
\end{eqnarray}

According to the Eq.(\ref{eq:R-fixed-length}), the expected codelength is an expression of $n$, which is the size of code alphabet of MZRL, and the count rate $q$. For a QKD system, $n$ is a parameter that should be adjusted carefully depending upon the requirements and available resources, while the count rate $q$ is almost constant. To analyze the optimization of the expected codelength, we need to confine the possible range of count rate. In general, the count rate $q$ is determined by the mean photon number $\mu $, the fibre loss coefficient $\alpha $, the distance $d$ between two parties, the inner loss ${\gamma _{Bob}}$ of the optical devices of Bob, the detection efficiency ${\eta _D}$ of Bob's detector, and the dark count rate $P_d$. The relationship is given by
\begin{eqnarray}
q = 1 - {e^{ - \mu  \cdot {{10}^{{{ - \left( {\alpha  \cdot d + {\gamma _{Bob}}} \right)} \mathord{\left/
 {\vphantom {{ - \left( {\alpha  \cdot d + {\gamma _{Bob}}} \right)} {10}}} \right.
 \kern-\nulldelimiterspace} {10}}}} \cdot {\eta _D}}} + {P_d}.
 \label{eq:count rate q}
 \end{eqnarray}
Table \ref{tbl-QKD Parameters} illustrates the typical value of these parameters above\cite{Xu2012phd}. To the best of our knowledge, the current maximal communication distance is about 250km\cite{stucki2009high,wang20122}, in which case the count rate $q$ is about $10^{-6}$. Besides, the count rates of most practical QKD systems are always less than 0.1\cite{sasaki2011field}. So it is reasonable to set the range of count rate $q$ as $\left[10^{-15}, 10^{-1}\right]$, which covers all possible values of current QKD systems.

\begin{table}[htbp]
\caption{The typical parameters related to the count rate of QKD.} \label{tbl-QKD Parameters}
\centering
\begin{tabular}{cccccc}
\hline
 {$\mu$} & {$\alpha $(dB/km)}& {$d$(km)} & {${\gamma _{Bob}}$(dB)} & {${\eta _D}$(\%)} & {$P_d$}\\
 \hline
0.5 & 0.2 & 0-250 & 4 & 10 & $10^{-5}$\\
\hline
\end{tabular}
\end{table}

Since the smaller $\overline{L}$ indicates the higher compression efficiency, our goal is to minimize the value of $\overline{L}$ under the constraint $q \in \left[10^{-15}, 10^{-1}\right]$. The optimization problem is hereby formalized in the Eq.(\ref{eq:optimal-problem-R(n)})

\begin{eqnarray}
\left\{ {\begin{array}{*{20}{l}}
{\mathop {\min }\limits_n }&{\overline{L}\left( n \right) = \frac{{q\left\lceil {{{\log }_2}n} \right\rceil }}{{1 - {{(1 - q)}^{n - 1}}}}}\\
{s.t.\;}&{n \in \mathbb{N}\;}\\
{}&{n \ge 2}\\
{}&{q \in \left[ {{{10}^{ - 15}},0.1} \right]}
\end{array}} \right.
\label{eq:optimal-problem-R(n)}
\end{eqnarray}

\subsection{optimization of the expected codelength}
For simplicity of expression, $g\left( n \right)$ is used to denote $1 - {(1 - q)^{n - 1}}$, then $\overline{L}\left( n \right)$ can be rewritten as
\[\overline{L}\left( n \right) = \frac{{q\left\lceil {{{\log }_2}n} \right\rceil }}{{g\left( n \right)}}.\]

It is easy to conclude that $g\left( n \right)$ is monotonic increasing function with respect to the variable $n$ for any given $0 < q < 1$, so
\[g\left( n \right) \le g\left( {{2^k}} \right),\;\;\forall n \in \left( {{2^{k - 1}},{2^k}} \right] \cap {\mathbb{N}^ + }, k \in {\mathbb{N}^ + }.\]
Besides, the value of the following expression
\[{q\left\lceil {{{\log }_2}n} \right\rceil }\]
is invariant in the range $n \in \left( {{2^{k - 1}},{2^k}} \right] \cap {\mathbb{N}^ + }$. So
\begin{eqnarray}
\overline{L}\left( n \right) \ge \overline{L}\left( {{2^k}} \right),\;\;\forall n \in \left( {{2^{k - 1}},{2^k}} \right] \cap {\mathbb{N}^ + }.
\label{eq:R(n)-R(k)}
\end{eqnarray}
Therefore we only need to consider the function values at ${2^k}$ and the optimization problem Eq.(\ref{eq:optimal-problem-R(n)}) is equivalent to
\begin{eqnarray}
\left\{ {\begin{array}{*{20}{l}}
{\mathop {\min }\limits_k }&{\overline{L}\left( k \right) = \frac{{qk}}{{1 - {{\left( {1 - q} \right)}^{{2^k} - 1}}}}}\\
{s.t.\;}&{k \in {\mathbb{N}^ + }\;}\\
{}&{q \in \left[ {{{10}^{ - 15}},0.1} \right]}
\end{array}} \right.
\label{eq:optimal-problem-R(k)}
\end{eqnarray}

To explore the properties of the function $\overline{L}\left( k \right)$ with respect to the variable $k$, the domain of $k$ is extended from $\mathbb{N}^+$ to real numbers no less than 1. That is
\begin{eqnarray}
\overline{L}(z) = \frac{{qz}}{{1 - {{(1 - q)}^{{2^z} - 1}}}},\;\;z \in \left[ {1, + \infty } \right).
\label{eq:L_z}
\end{eqnarray}

\begin{thm}
For any given $q \in \left(0,0.1\right]$, there exists a constant ${z_0} \in \left( { - {{\log }_2}\left( { - \ln \left( {1 - q} \right)} \right), + \infty } \right)$ satisfying that the function $\overline{L}\left( z \right)$ monotonically decreases in the domain $z \in \left[ {1,{z_0}} \right)$, monotonically increases in the domain $z \in \left( {{z_0}, + \infty } \right)$, and reaches the global minimum at the point $z=z_0$. In other words,
\begin{equation}
\left\{ {\begin{array}{*{20}{l}}
{\frac{{\partial \overline{L}}}{{\partial z}} < 0,}&{{\rm{when}}\;z \in \left[ {1,{z_0}} \right)}\\[1mm]
{\frac{{\partial \overline{L}}}{{\partial z}} = 0,}&{{\rm{when}}\;z = {z_0}}\\[1mm]
{\frac{{\partial \overline{L}}}{{\partial z}} > 0,}&{{\rm{when}}\;z \in \left( {{z_0}, + \infty } \right)}
\end{array}} \right..
\label{eq:partial_derivative_R_z}
\end{equation}
\label{Thm-R property}
\end{thm}
\begin{IEEEproof}
Please refer to Appendix ~\ref{sec:Appendixes Proof of Lemma}.
\end{IEEEproof}

\begin{figure}[htbp]
\includegraphics[width=0.5\textwidth]{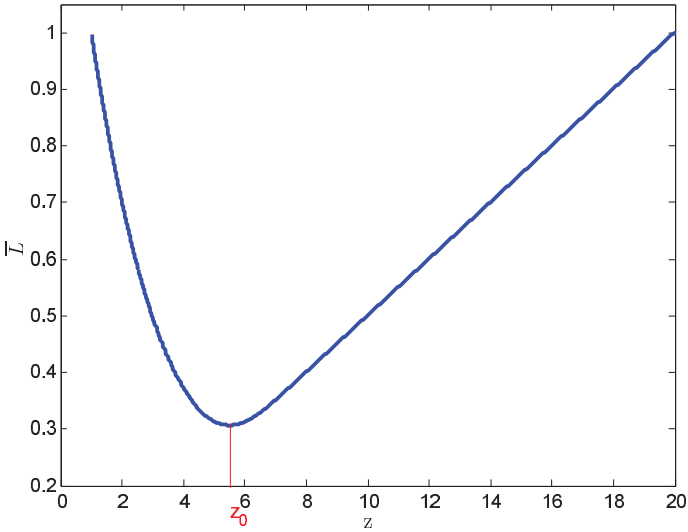}
\caption{\label{fig:Theory4_example} The illustration of Theorem \ref{Thm-R property} in which case $q=0.05$.}
\end{figure}

An example curve of $\overline{L}\left(z\right)$ demonstrating the Theorem \ref{Thm-R property} is shown as Fig. \ref{fig:Theory4_example}, where $q=0.05$.

\begin{lem}
The optimal solution of Eq.(\ref{eq:optimal-problem-R(k)}) is reached at $z=\left\lfloor {{z_0}} \right\rfloor$ or $z=\left\lceil {{z_0}} \right\rceil$.
\label{lem_R_k_k0}
\end{lem}

Since $k \in \mathbb{N}^+$, the Lemma \ref{lem_R_k_k0} is a straightforward derivation of the Theorem \ref{Thm-R property}.

So far, the existence of the optimal solution of Eq.(\ref{eq:optimal-problem-R(k)}) has been proved, and the optimal parameter $k_{opt}$ has been determined as $\left\lfloor {{z_0}} \right\rfloor$ or $\left\lceil {{z_0}} \right\rceil$. So the next task is to solve the key value $z_0$.

\begin{thm}\label{Thm-Bound of z0}
For any given $q \in \left[10^{-15}, 0.1\right]$, the point $z_0$ which leads to the global minimum of function $\overline{L}\left(z\right)$ is bounded by
\[ - {\log _2}\left( { - \ln \left( {{\rm{1 - q}}} \right)} \right) < {z_0} <  - {\log _2}\left( { - \ln \left( {{\rm{1 - q}}} \right)} \right) + 3.\]
\end{thm}
\begin{IEEEproof} According to the Theorem \ref{Thm-R property}, we have
\begin{eqnarray}
{z_0} >  - {\log _2}\left( { - \ln \left( {{\rm{1 - q}}} \right)} \right),\forall q \in \left[ {{{10}^{ - 15}},0.1} \right].
\label{eq:z0_lower_bound}
\end{eqnarray}
At the same time, the value of the partial deviation $\frac{{\partial \overline{L}}}{{\partial z}}$ at $z =  - {\log _2}\left( { - \ln \left( {{\rm{1 - q}}} \right)} \right) + 3$ is given by
\begin{eqnarray}
\frac{{\partial \overline{L}}}{{\partial z}}{|_{z =  - {{\log }_2}\left( { - \ln \left( {1 - q} \right)} \right) + 3}} = \frac{{{e^8}(1 - q)q}}{{{{\left( {1 - {e^8}\left( {1 - q} \right)} \right)}^2}}}A\left( q \right),
\label{eq:R_partial_k_at_+3}
\end{eqnarray}
where
\[A\left( q \right) =  - 1 - 24\ln 2 + {e^8}(1 - q) + 8\ln \left( { - \ln \left( {1 - q} \right)} \right).\]
Since
\begin{eqnarray}
\frac{{{e^8}(1 - q)q}}{{{{\left( {1 - {e^8}\left( {1 - q} \right)} \right)}^2}}} > 0, \forall q \in \left[10^{-15}, 0.1\right]
\label{eq:R_partial_k_at_+3 -15-part1}
\end{eqnarray}
the sign of Eq.(\ref{eq:R_partial_k_at_+3}) is same as the sign of $A\left( q \right)$. The partial derivative of $A\left( q \right)$ can be evaluated as
\[\frac{{\partial A}}{{\partial q}} =  - {e^8} - \frac{8}{{(1 - q)\ln \left( {1 - q} \right)}},\]
which is a monotonic decreasing function with respect to the variable $q$ and
\[\left\{ {\begin{array}{*{20}{l}}
{\frac{{\partial A}}{{\partial q}} > 7.20 \times {{10}^{15}},}&{when\;q = {{10}^{ - 15}}}\\[1mm]
{\frac{{\partial A}}{{\partial q}} <  - 2.80 \times {{10}^3},}&{when\;q = 0.1}
\end{array}} \right.\]
So the function $A\left( q \right)$ is firstly monotonic increasing and then monotonic decreasing in the range $q \in \left[10^{-15}, 0.1\right]$. The minimum must occur at the point $q=10^{-15}$ (${A\left( {{{10}^{ - 15}}} \right) \approx 2687.85}$) or $q=0.1$ (${A\left( {0.1} \right) \approx {\rm{2647}}{\rm{.22}}}$), so
\begin{eqnarray}
A\left( q \right) > 0,\forall q \in \left[10^{-15}, 0.1\right].
\label{eq:R_partial_k_at_+3 -15-part2}
\end{eqnarray}
Hence, combining the Eq.(\ref{eq:R_partial_k_at_+3}), Eq.(\ref{eq:R_partial_k_at_+3 -15-part1}) and Eq.(\ref{eq:R_partial_k_at_+3 -15-part2}),
\begin{eqnarray}
\frac{{\partial \overline{L}}}{{\partial z}}{|_{z =  - {{\log }_2}\left( { - \ln \left( {1 - q} \right)} \right) + 3}} > 0,\forall q \in \left[10^{-15}, 0.1\right].
\label{eq:R_partial_k_at_+3 -15}
\end{eqnarray}
Making use of Eq.(\ref{eq:R_partial_k_at_+3 -15}) and the Theorem \ref{Thm-R property}, it can be concluded that
\begin{eqnarray}
{z_0} <  - {\log _2}\left( { - \ln \left( {1 - q} \right)} \right) + 3,\forall q \in \left[ {{{10}^{ - 15}},0.1} \right].
\label{eq:z0_upper_bound}
\end{eqnarray}
Combining the Eq.(\ref{eq:z0_lower_bound}) and Eq.(\ref{eq:z0_upper_bound}), the theorem is proved.
\end{IEEEproof}

According to the Lemma \ref{lem_R_k_k0} and the Theorem \ref{Thm-Bound of z0}, the optimal parameter $k_{opt}$ is one of the following five values, i.e. $\left\lfloor y \right\rfloor ,\left\lfloor y \right\rfloor  + 1,\left\lfloor y \right\rfloor  + 2,\left\lfloor y \right\rfloor  + 3,\left\lceil y \right\rceil  + 3$, where $y$ is the brief denotation of $ - {\log _2}\left( { - \ln \left( {1 - q} \right)} \right)$. Subsequently, an efficient iterative solution of Eq.(\ref{eq:optimal-problem-R(k)}) is presented, which is called Algorithm \ref{Algo:1}.

\setcounter{algleo}{0}
\begin{algleo}
\linonumber \AlgorithmFlag \label{Algo:1}  The solution of optimization problem Eq.(\ref{eq:optimal-problem-R(k)}).
\linonumber \textbf{Input:} Count rate $q$.
\linonumber \textbf{Output:} The optimal solution $\overline{L}_{opt}$, and the optimal parameter $k_{opt}$.
\li $k = \left\lfloor { - {{\log }_2}\left( { - \ln \left( {1 - q} \right)} \right)} \right\rfloor$.
\li $\overline{L}_1 = \overline{L}\left( k \right)$.
\li \label{algo1_Iter_begin}{\bf while}  {(1)} {\bf do}
\begin{algleo}
    \li $k=k+1$.
    \li $\overline{L}_2=\overline{L}\left( k \right)$.
    \li {\bf if} $\left(\overline{L}_1 \le \overline{L}_2\right)$ {\bf then}
    \begin{algleo}
        \li break.
    \end{algleo}
    \li {\bf else}
    \begin{algleo}
        \li $\overline{L}_1=\overline{L}_2$.
    \end{algleo}
    \li {\bf end if}
\end{algleo}
\li \label{algo1_Iter_End}{\bf end while}
\li $\overline{L}_{opt}=\overline{L}_1$.
\li $k_{opt}=k-1$.
\end{algleo}

\begin{figure}[htbp]
\includegraphics[width=0.5\textwidth]{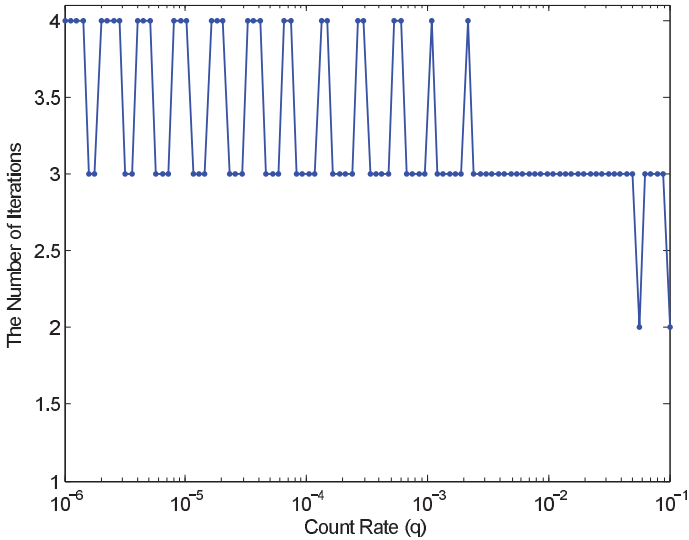}
\caption{\label{fig:iteration_numbers} The number of iterations of Algorithm \ref{Algo:1} as a function of the count rate $q$.}
\end{figure}

For convenience, the steps \ref{algo1_Iter_begin} - \ref{algo1_Iter_End} of Algorithm \ref{Algo:1} are called one iteration. It is obvious that the algorithm converges within five iterations for any given $q \in \left[10^{-15}, 0.1\right]$. Some count rates $q$ between $10^{-6}$ and $10^{-1}$ are chosen as the input of Algorithm \ref{Algo:1}, and the corresponding numbers of iterations are demonstrated in Fig. \ref{fig:iteration_numbers}. The experimental results show that the largest number of iterations is 4 and the average number of iterations is 3.28.

To sum up, the optimal solution of Eq.(\ref{eq:optimal-problem-R(n)}) stated in the previous section is solved when the size of codeword alphabet $n=2^{k_{opt}}$.

\subsection{\label{performance analysis}theoretical performance analysis}
\subsubsection{\label{performance analysis-compression efficiency}compression efficiency}
\begin{figure}[htbp]
\includegraphics[width=0.5\textwidth]{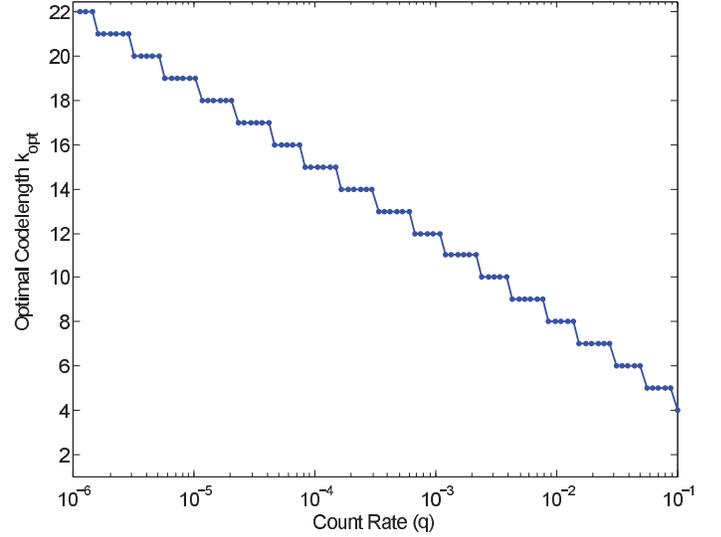}
\caption{\label{fig_k_opt} The optimal codelength of $S$ $k_{opt}\left(q\right)$.}
\end{figure}

\begin{figure}[htbp]
\includegraphics[width=0.5\textwidth]{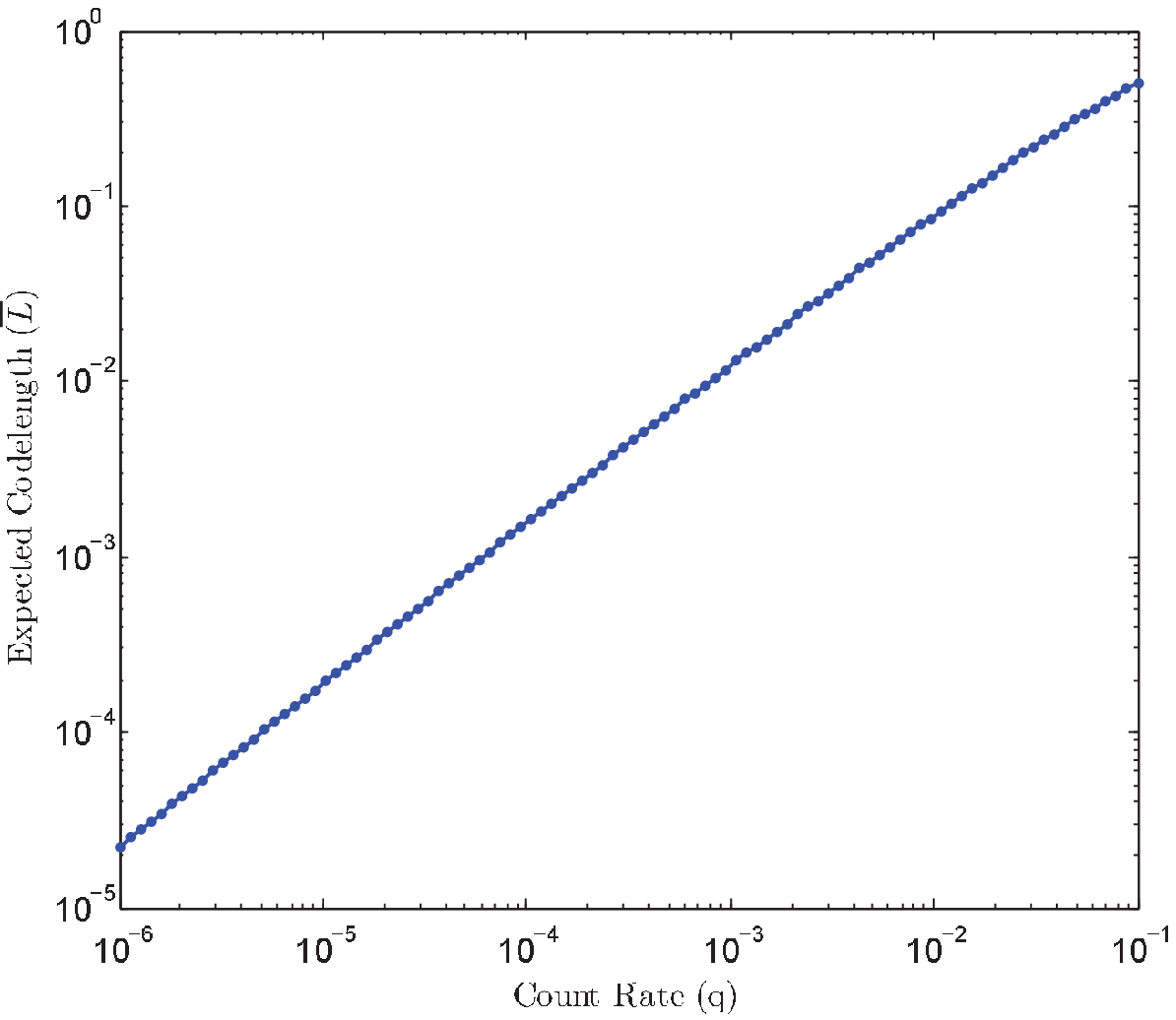}
\caption{\label{fig_expected_codelength} The expected codelength $\overline{L}\left(q\right)$.}
\end{figure}

\begin{figure}[htbp]
\includegraphics[width=0.5\textwidth]{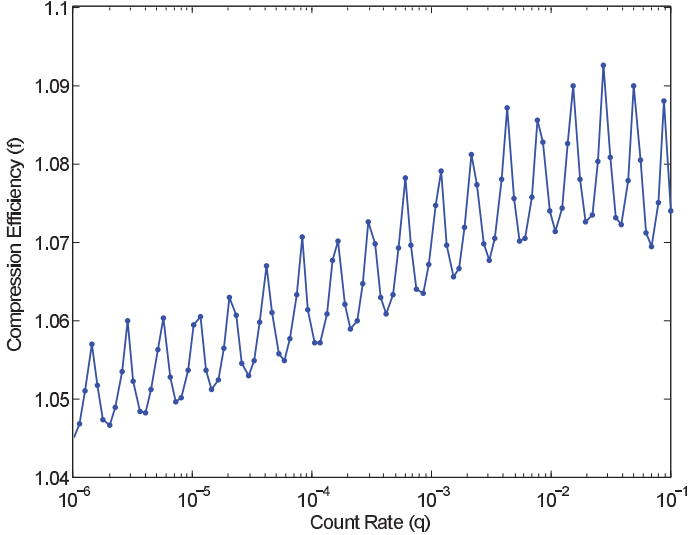}
\caption{\label{fig_Compression_efficiency} The compression efficiency $f\left(q\right)$.}
\end{figure}

To compute the compression efficiency $f = \frac{{\overline L }}{{H(X)}}$, first of all we need to calculate the expected codelength $\overline L$. One hundred different count rates are selected in the range $\left[10^{-6}, 10^{-1}\right]$ on the logarithmic scale. The optimal codelength of n-ary source $S$, i.e. $k_{opt}$, is computed via Algorithm \ref{Algo:1} for each count rate and the corresponding optimal size of codeword alphabet $n_{opt}$ is $2^{k_{opt}}$. The expected codelength $\overline L (n)$ can be hereby computed according to Eq.(\ref{eq:R-fixed-length}). The entropy of binary source can be obtained by straightforward application of Definition \ref{defn:Entropy}, i.e. $H\left( X \right) =  - q{\log _2}q - \left( {1 - q} \right){\log _2}\left( {1 - q} \right)$, denoted as $h\left( q \right)$. So far, the compression efficiency $f = \frac{{\overline L }}{{h\left( q \right)}}$ can be obtained. The theoretical results of $k_{opt}$, $\overline{L}$ and $f$ are shown as Fig \ref{fig_k_opt} - \ref{fig_Compression_efficiency} respectively.

The compression efficiency $f$ is less than 1.10 during the whole domain of count rate. So it is demonstrated that the compression performance of our MZRL source coding is very close to the Shannon's limit.

\subsubsection{\label{sec:time complexity}time complexity}
In this section, we will analyse the time complexity of proposed bit sifting scheme in Fig.\ref{fig:bit_sifting_schematic}.

Bit sifting at Bob side consists of the \emph{Undetected Bit Removal} and the \emph{Source Encoder}. For each $\rm{original\_key}_B$, the \emph{Undetected Bit Removal} determines whether it is a valid detection or not, and outputs the detected validity to the \emph{Source Encoder}. The \emph{Source Parse} of the \emph{Source Encoder} in Fig.\ref{fig:MZRL coding scheme} determines whether a valid message $s_i$ is formed, i.e. the current input detected validity is "1" or the current zero counter $i$ is equal to $n-1$. If not, $i = i+1$. Otherwise, the message $s_i$ is output to the \emph{Message Encoder} which outputs the corresponding codeword $c_i$, and the zero counter $i$ is reset to 0. Since the time complexity of both the \emph{Undetected Bit Removal} and the \emph{Source Encoder} are constant, the time complexity of bit sifting for each $\rm{original\_key}_B$ is also constant. Hence, when $m$ $\rm{original\_key}_B$ are input, the time complexity of bit sifting at Bob side is $O\left(m\right)$.

Once receiving a codeword $c_i$, the \emph{Source Decoder} at Alice side in Fig.\ref{fig:bit_sifting_schematic}, whose time complexity is constant, decodes it to the corresponding message $s_i$, and outputs the $s_i$ to the \emph{Undetected Bit Removal}. If $i$ is equal to $n-1$, the \emph{Undetected Bit Removal} discards $n-1$ consecutive $\rm{original\_key}_A$ from the \emph{Buffer}. Otherwise the \emph{Undetected Bit Removal} discards the former $i$ $\rm{original\_key}_A$ and reserve the $i+1\text{th}$ as a $\rm{raw\_key}_A$. Therefore the time complexity of bit sifting for the input codeword $c_i$ is $O\left(i\right)$. Assuming that the received $w$ codewords are $c_{i_0}$, $c_{i_1}$, $\cdots$, $c_{i_{w-1}}$, the time complexity of bit sifting is
\[O\left( {\sum\limits_{j = 0}^{w-1} {{i_j}} } \right).\]
In fact, ${\sum\limits_{j = 0}^{w-1} {{i_j}} }$ is the number of processed $\rm{original\_key}_A$ so the time complexity of bit sifting at Alice side is also $O\left(m\right)$.

In summary, the time complexity of the bit sifting at both Alice and Bob sides are linearly dependent on the number of original keys $m$.

\subsubsection{space complexity}
In this section, we will analyse the space complexity of proposed bit sifting scheme in Fig.\ref{fig:bit_sifting_schematic}. Here, we assume that Bob does not cache more than one codeword but send a codeword to Alice as soon as it is formed. In this case, the bit sifting at Bob side only need to store two temporary variables, i.e. one zero counter and one codeword. Both of them are represented by $\left\lceil {{{\log }_2}n} \right\rceil$ bits, so the space complexity of bit sifting at Bob side is $O\left({{{\log }_2}n}\right)$.

Since Alice has to store the $\rm{original\_key}_A$ in the \emph{Buffer} till she receives a codeword carrying the detected validity of the $\rm{original\_key}_A$ from Bob, the required storage consists of some temporary variables and the \emph{Buffer} used to store $\rm{original\_key}_A$. The temporary variables are the received codeword $c_i$ and the message $s_i$ which are both represented by $\left\lceil {{{\log }_2}n} \right\rceil$ bits. While the size of the \emph{Buffer} depends on the time difference $t_{\rm{diff}}$ between the time when an $\rm{original\_key}_A$ is stored in the \emph{Buffer} and the time it is removed from the \emph{Buffer} by the \emph{Undetected Bit Removal}. According to the MZRL codes, Alice has to send Bob at most $n-1$ qubits to form a codeword. So the maximum time difference is
\begin{equation}
{t_{\rm{diff}}} = \left( {n - 2} \right){t_{rf}} + {t_2} + {t_3} + {t_4} + {t_5}.
\end{equation}
$\left( {n - 2} \right){t_{rf}}$ means the time that Alice prepares $n-1$ qubits, where $t_{rf}$ is the reciprocal of the repetition frequency of QKD.\\
$t_2$ is the time that the ${n-1}th$ qubit is transmitted from Alice to Bob over quantum channel, which depends on the distance $d$ between Alice and Bob.\\
$t_3$ is the time that the bit sifting at Bob side processes the ${n-1}th$ $\rm{original\_key}_B$, which is a constant according to the analysis in section \ref{sec:time complexity}. By then, the codeword $c_{n-2}$ or $c_{n-1}$ is formed.\\
$t_4$ is the time that the codeword is transmitted from Bob to Alice over authenticated classical channel, which also depends on the distance $d$. \\
$t_5$ is the time that the \emph{Source Decoder} at Alice side decodes the codeword to the corresponding message, which is also a constant according to the analysis in section \ref{sec:time complexity}. So far, the \emph{Undetected Bit Removal} can begin to discard these $n-1$ $\rm{original\_key}_A$ from the \emph{Buffer}. \\
Hence the number of the cached $\rm{original\_key}_A$ in the \emph{Buffer} is
\[\frac{{{t_{diff}}}}{{{t_{rf}}}} + 1 = \left( {n - 1} \right) + \frac{{{t_2} + {t_3} + {t_4} + {t_5}}}{{{t_{rf}}}},\]
which is $O\left(n\right)$. Therefore, the space complexity of bit sifting at Alice side is $O\left(n\right)$.

Since the size of code alphabet $n$ exponentially dependents on the optimal parameter $k_{opt}$, the required storage at Alice side may be very large. For instance, let $q=10^{-6}$, then the optimal parameter $k_{opt}$ is 22 according to Algorithm \ref{Algo:1}, and the required storage at Alice side is about multiple times 4Mb. The times depend on the protocol of the QKD systems.  In fact, memory resource sometimes may be very expensive, such as FPGA based QKD system\cite{stucki2009high,zhang2012real,walenta2014fast}. Although the storage of FPGA can be extended by attaching several SRAMs or DDRs, the performance of SRAM or DDR is not as good as the inner storage of FPGA. In the case of limited storage resource, the optimization problem Eq.(\ref{eq:optimal-problem-R(n)}) can be rewritten as
\begin{eqnarray}
\left\{ {\begin{array}{*{20}{l}}
{\mathop {\min }\limits_n }&{\overline{L}\left( n \right) = \frac{{q\left\lceil {{{\log }_2}n} \right\rceil }}{{1 - {{(1 - q)}^{n - 1}}}}}\\
{s.t.\;{\kern 1pt} }&{n \in {\mathbb{N} }\;{\kern 1pt} }\\
{}& { {n_{\max }} \ge n \ge 2 }\\[1mm]
{}& { q \in \left[ {10^{-15},0.1} \right] }
\end{array}} \right.
\label{eq:optimal-R(k)-finite-resource}
\end{eqnarray}
, where $n_{max}$ is the possible maximal code alphabet size, which can be evaluated according to the available storage size. The optimal solution of Eq.(\ref{eq:optimal-R(k)-finite-resource}) is stated in the Theorem \ref{Thm-optimization solution of limited case}.

\begin{thm}\label{Thm-optimization solution of limited case}
For any given $q \in \left[10^{-15}, 0.1\right]$ and $n_{max}$, if $n_{\max} \ge 2^{k_{opt}}$ then the optimal solution of Eq.(\ref{eq:optimal-R(k)-finite-resource}) is reached at $n=2^{k_{opt}}$. Otherwise it is reached at $n = {2^{\left\lfloor {{{\log }_2}{n_{\max }}} \right\rfloor }}$ or $n=n_{max}$.
\end{thm}
\begin{IEEEproof}
For brevity, let $\left\{ {a...b} \right\} \buildrel \Delta \over = \left[ {a,b} \right] \cap \mathbb{N}.$

\textbf{(a)} When $n_{\max} \ge 2^{k_{opt}}$, the optimal code alphabet size
\[{n_{opt}} = {2^{{k_{opt}}}} \le {n_{\max }},\]
which is reachable in the domain $\left\{ {2...{n_{\max }}} \right\}$. So the optimal solution of Eq.(\ref{eq:optimal-R(k)-finite-resource}) is reached at $n=2^{k_{opt}}$ in this case.

\textbf{(b)} When $n_{\max} < 2^{k_{opt}}$, the optimal code alphabet size
\[{n_{opt}} = {2^{{k_{opt}}}} > {n_{\max }},\]
which cannot be reachable in the domain $\left\{ {2...{n_{\max }}} \right\}$. In the case, the domain $\left\{ {2...{n_{\max }}} \right\}$ is divided into $\left\{ {2...{2^{{k_{\max }}}}} \right\}$ and $\left\{ {{2^{{k_{\max }}}} + 1...{n_{\max }}} \right\}$, where ${k_{\max }} = \left\lfloor {{{\log }_2}{n_{\max }}} \right\rfloor$.\\[1mm]

\textbf{(b.1)} Since
\[\left\{ {2...{2^{{k_{\max }}}}} \right\} = \bigcup\limits_{k = 1}^{{k_{\max }}} {\left\{ {{2^{k - 1}}+1...{2^k}} \right\}} \]
and the minimum of $\overline{L}\left(n\right)$ in the range ${\left\{ {{2^{k - 1}}+1...{2^k}} \right\}}$ occurs at the point $2^k$ according to Eq.(\ref{eq:R(n)-R(k)}), $1 \le k \le k_{max}$, we just need to consider the minimum of $\overline{L}\left(k\right)$ in the range $\left\{ {1...{k_{\max }}} \right\}$. According to Lemma \ref{lem_R_k_k0}, it can be inferred that $k_{opt} \le \left\lceil {{z_0}} \right\rceil$ and
\[{k_{\max }}  = \left\lfloor {{{\log }_2}{n_{\max }}} \right\rfloor \le {k_{opt}} - 1 \le \left\lceil {{z_0}} \right\rceil  - 1 < {z_0}.\]
So according to Theorem \ref{Thm-R property}, $\overline{L}\left(k\right)$ is monotonic decreasing in the range $\left\{ {1...{k_{\max }}} \right\}$ and reaches the minimum at the point $k=k_{max}$. Therefore, the minimum of $\overline{L}\left(n\right)$ in the range $\left\{ {2...{2^{{k_{\max }}}}} \right\}$ occurs at the point $n=2^{k_{max}}$, i.e. $n=2^{\left\lfloor {{{\log }_2}{n_{\max }}} \right\rfloor}$.

\textbf{(b.2)} Since the function $\overline{L}\left(n\right)$ is monotonic decreasing in the range $\left\{ {{2^{{k_{\max }}}} + 1...{n_{\max }}} \right\}$, the minimum of $\overline{L}\left(n\right)$ in the range occurs at the point $n={n_{\max }}$.

Combining \textbf{(b.1)} and \textbf{(b.2)}, it can be seen that the optimal solution of Eq.(\ref{eq:optimal-R(k)-finite-resource}) is reached at $n=2^{\left\lfloor {{{\log }_2}{n_{\max }}} \right\rfloor}$ or $n=n_{max}$ when $n_{\max} < 2^{k_{opt}}$.

Combining the case \textbf{(a)} and the case \textbf{(b)}, the theorem is proved.
\end{IEEEproof}

\begin{figure}[htbp]
\includegraphics[width=0.5\textwidth]{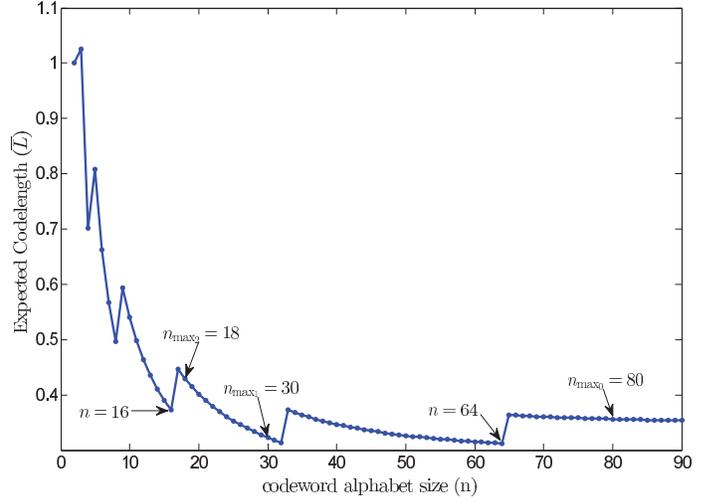}
\caption{\label{fig_Rn_piecewise_monotonic} The illustration of the Theorem \ref{Thm-optimization solution of limited case} where $q=0.05$.}
\end{figure}

Figure \ref{fig_Rn_piecewise_monotonic} shows an example curve of $\overline{L}\left(n\right)$ with three preset $n_{\max}$, which demonstrates the different aspects of Theorem \ref{Thm-optimization solution of limited case}. Here the count rate $q=0.05$, $k_{opt}$ is 6 according to Algorithm \ref{Algo:1}, and the optimal solution is reached at
\[n = \left\{ {\begin{array}{*{20}{l}}
{{2^{{k_{opt}}}} = 64,}&{when\;{n_{\max }} = {n_{{\rm{ma}}{{\rm{x}}_0}}} = 80}\\
{{n_{\max }} = 30,}&{when\;{n_{\max }} = {n_{{\rm{ma}}{{\rm{x}}_1}}} = 30}\\[1mm]
{{2^{\left\lfloor {{{\log }_2}{n_{\max }}} \right\rfloor }} = 16,}&{when\;{n_{\max }} = {n_{{\rm{ma}}{{\rm{x}}_2}}} = 18}
\end{array}} \right.\]

According to the Theorem \ref{Thm-optimization solution of limited case}, Algorithm \ref{Algo:2} is presented to obtain the optimal solution of Eq.(\ref{eq:optimal-R(k)-finite-resource}). Its convergence can be deduced directly from the convergence of Algorithm \ref{Algo:1}.

\setcounter{algleo}{0}
\begin{algleo}
\linonumber \AlgorithmFlag \label{Algo:2}  The solution of optimization problem Eq.(\ref{eq:optimal-R(k)-finite-resource}).
\linonumber \textbf{Input:} Count rate $q$, and possible maximal code alphabet size $n_{max}$.
\linonumber \textbf{Output:} The optimal solution $\overline{L}_{opt}$, and the optimal parameter $n_{opt}$.
\li $k_{max}=\left\lfloor {{{\log }_2}{n_{\max }}} \right\rfloor$.
\li Let $q$ be the input of Algorithm \ref{Algo:1}, then $\overline{L}_{opt}$ and $k_{opt}$ can be obtained.
\li {\bf if} $\left(n_{max} \ge 2^{k_{opt}}\right)$ {\bf then}
\begin{algleo}
    \li $n_{opt}=2^{k_{opt}}$.
\end{algleo}
\li {\bf else if} $\overline{L}\left(2^{k_{max}}\right) > \overline{L}\left(n_{max}\right)$
\begin{algleo}
    \li $n_{opt}=n_{max}$.
    \li $\overline{L}_{opt}=\overline{L}\left(n_{max}\right)$.
\end{algleo}
\li {\bf else}
\begin{algleo}
    \li $n_{opt}=2^{k_{max}}$.
    \li $\overline{L}_{opt}=\overline{L}\left(2^{k_{max}}\right)$.
\end{algleo}
\li {\bf end if}
\end{algleo}

\section{\label{sec:Experimental-Results}Experimental Results and Analysis}
\subsection{compression efficiency}
In the experiment, one hundred different count rates are selected in the range $\left[10^{-6}, 10^{-1}\right]$ on the logarithmic scale and the simulation results are obtained by processing $10^{10}$ original keys for each count rate.

\begin{figure}[b]
\includegraphics[width=0.5\textwidth]{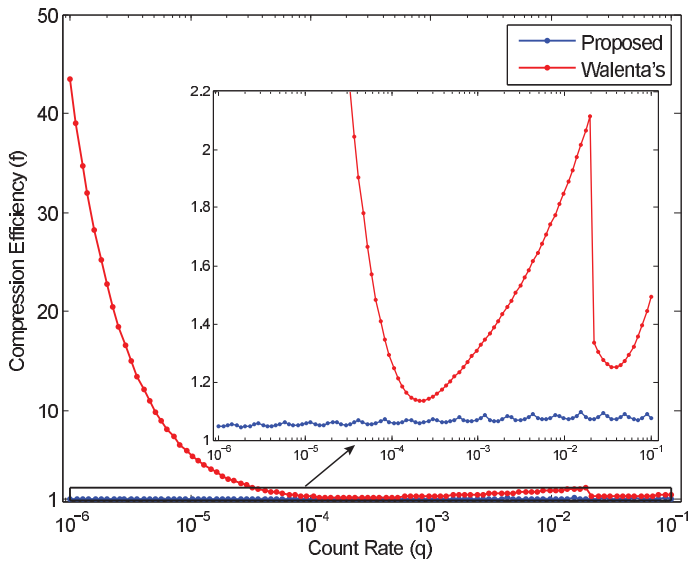}
\caption{\label{fig_Compare_with_walenta} The comparison of compression efficiency $f$ between the proposed scheme and Walenta's scheme \cite{walenta2014fast}. }
\end{figure}

Figure \ref{fig_Compare_with_walenta} demonstrates the compression efficiency $f$ of the proposed bit sifting scheme and the bit sifting scheme of \cite{walenta2014fast}. The performance of the scheme of \cite{walenta2014fast} is near the Shannon limit for $q \in \left[ {{{10}^{ - 4}},{{10}^{ - 1}}} \right]$, while falls sharply as the count rate is outside of the range. It is clear that the compression efficiency of our scheme is always near the Shannon limit and superior to the scheme of \cite{walenta2014fast} in the whole range of the count rate.

\subsection{secure key consumption}
In \cite{walenta2014fast}, Walenta etc. use a combination\cite{stinson1994universal} of ${\varepsilon} \text{-almost}$ strongly universal hash functions and a family of strongly universal hash functions named polynomial hashing\cite{Wegman1979Universal,Wegman1981New} to achieve information theoretically secure authentication. The authentication algorithm produces a 127-bit authentication tag for every $2^{20}$ bits of classical communication, and consumes 383 secure keys to select a hash function for every tag. According to the result of \cite{portmann2012key}, the same hash function can be reused for multiple authentication rounds if the tags attached to the messages are one-time pad encrypted, so only 127 secure keys are consumed for the classical communication of every $2^{20}$ bits and the key consumption can be reduced to one third. Although the authentication scheme is very efficient, the key consumptions are still 2.7\% and 5\% of the generated secure key of the QKD system when the fibre length is 1km and 25km respectively.

Let $M$ be the amount of classical communication, then the key consumption is
\[K = 127 \cdot \left\lceil {\frac{M}{{{2^{20}}}}} \right\rceil. \]
Especially, the key consumption for the bit sifting is
\[{K_{bs}} = 127 \cdot \left\lceil {\frac{{m \cdot h\left( q \right) \cdot f}}{{{2^{20}}}}} \right\rceil, \]
where $m$ is the number of original keys to be processed, $q$ is the count rate, and $f$ is the compression efficiency. Since \[\frac{{m \cdot h\left( q \right)}}{{{2^{20}}}}f \le \left\lceil {\frac{{m \cdot h\left( q \right) \cdot f}}{{{2^{20}}}}} \right\rceil  < \frac{{m \cdot h\left( q \right)}}{{{2^{20}}}}f + 1\] and $\frac{{m \cdot h\left( q \right)}}{{{2^{20}}}}f$ is usually very large because $f \ge 1$ and QKD is a continuous high speed system which leads to the large $\frac{{m \cdot h\left( q \right)}}{{{2^{20}}}}$, we have \[\left\lceil {\frac{{m \cdot h\left( q \right) \cdot f}}{{{2^{20}}}}} \right\rceil  \approx \frac{{m \cdot h\left( q \right)}}{{{2^{20}}}}f.\]
So
\begin{eqnarray}
\frac{{{K_{bs - A}}}}{{{K_{bs - B}}}} \approx \frac{{{f_A}}}{{{f_B}}},
\label{eq:K-bs-A-K-bs-B}
\end{eqnarray}
where subscript $A$ and $B$ indicate two different bit sifting schemes. The Eq.(\ref{eq:K-bs-A-K-bs-B}) demonstrates that the key consumption depends linearly on the compression efficiency. Since the compression efficiency of proposed scheme is always superior to that of the scheme of \cite{walenta2014fast}, the key consumption of proposed scheme is always less.

More experiment results of the proposed scheme and the Walenta's scheme are listed in Table \ref{tbl-count rate of walenta}, including compression efficiency and the ratio of the two compression efficiencies. It needs to explain that the count rate is not given explicit in \cite{walenta2014fast}, while it can be inferred according to the sifted key rate and the repetition frequency of the QKD system. It can be seen that if the QKD system in \cite{walenta2014fast} employs our proposed scheme for bit sifting, 29\%, 19\% and 17\% of the secure key consumption for bit sifting can be saved when the fibre length is 1km, 12.5km and 25km respectively. Besides, more than 88\% of the key consumption of the post-processing comes from the bit sifting step, which is evaluated according to the sifting scheme and the communication rate among the procedures of post-processing presented in \cite{walenta2014fast}. So our scheme can greatly save the secure key consumption of the whole QKD post-processing system.
\begin{table}[htbp]
\caption{The comparison between the performances of proposed scheme and Walenta's scheme \cite{walenta2014fast} for the QKD system in  \cite{walenta2014fast}. The subscript $w$ indicates Walenta's scheme, while the subscript $p$ indicates the proposed scheme.} \label{tbl-count rate of walenta}
\centering
\begin{tabular}{ccccc}
\hline
Fibre Length & Count Rate & $f_w$ &   $f_p$ & ${{{f_p}} \mathord{\left/
 {\vphantom {{{f_p}} {{f_w}}}} \right.\kern-\nulldelimiterspace} {{f_w}}}$ \\
\hline
1 km &   $2.76*10^{-3}$  &  1.51   & 1.07 &  0.71 \\
12.5 km & $1.18*10^{-3}$  & 1.34  & 1.08 & 0.81 \\
25 km &  $7.87*10^{-4}$  &  1.27   & 1.06 &  0.83 \\
\hline
\end{tabular}
\end{table}

\subsection{some suggestions for some representative QKD systems}

\begin{table*}
\caption{The parameters of four QKD systems.} \label{tbl-QKD system detail arguments}
\centering
\begin{threeparttable}
\begin{tabular}{lccccc}
\hline
QKD System & Dark Count Rate $P_d$ & Mean Photon Number $\mu$ & Distance $d$(km) &       Loss(dB){\color{blue}\tnote{a}} & Detection Efficiency ${\eta _D}$ \\
\hline
Dixon etc.\cite{dixon2008gigahertz} &   $1.36*10^{-5}$ &       0.55 &      20 &      8.01  &     10.00\% \\
Stucki etc.\cite{stucki2009high} &   $1.60*10^{-8}$ &        0.50 &        250 &     42.60  &      2.65\% \\
Zhang etc.\cite{zhang2012real} &   $1.00*10^{-5}$ &        0.60 &         20 &      7.20  &     12.00\% \\
Tanaka etc.\cite{tanaka2012high} &   $2.00*10^{-5}$ &        0.40 &         50 &     14.00  &     10.00\% \\
\hline
\end{tabular}
\begin{tablenotes}
        \footnotesize
        \item[a] The parameter includes both the loss of transmission and the inner loss ${\gamma _{Bob}}$ of Bob's optical devices.
\end{tablenotes}
\end{threeparttable}
\end{table*}

\begin{table*}
\caption{The recommendations for four representative QKD systems.} \label{tbl-QKD system recommended k}
\centering
\begin{threeparttable}
\begin{tabular}{lcccccc}
\hline
QKD System & Count Rate $q$ & Theoretical $n$ & Theoretical $f$ & Available Storage & Recommended $n$ & Actual $f$\\
\hline
Dixon etc.\cite{dixon2008gigahertz} & $8.68*10^{-3}$ &          $2^8$ &  1.08 &      $\ge 2$GB{\color{blue}\tnote{a}} &          $2^8$ &    1.08  \\
Stucki etc.\cite{stucki2009high} & $7.44*10^{-7}$ &         $2^{22}$ & 1.06 &       32Kb{\color{blue}\tnote{b}} &         $12*2^{10}$ &      70.57{\color{blue}\tnote{c}} \\
Zhang etc.\cite{zhang2012real} & $1.36*10^{-2}$ &          $2^8$ & 1.08 &     32Mb{\color{blue}\tnote{d}} &          $2^8$ &    1.08  \\
Tanaka etc.\cite{tanaka2012high} & $1.42*10^{-3}$ &         $2^{11}$ & 1.07 &       $ 833$MB{\color{blue}\tnote{e}} &         $2^{11}$ &     1.07 \\
\hline
\end{tabular}
\begin{tablenotes}
\footnotesize
    \item[a] The system is implemented in PC and the available storage is estimated to be larger than 2GB.
    \item[b] The system is implemented in Virtex II Pro FPGA, and 32Kb memory for sifting.
	\item[c] If the storage for sifting is extended to 8Mb, the actual compression efficiency would be 1.06.
	\item[d] The system is implemented in two Cyclone III series FPGAs (EP3C120), and 32Mb memory for sifting.
	\item[e] The system is implemented in Several FPGAs, and average 833MB memory is used for each FPGA.
\end{tablenotes}
\end{threeparttable}
\end{table*}

Many QKD systems have been developed since the first QKD system was developed in 1984. Most of them failed to take into account the authentication of the classical channel, so they didn't pay much attention to the communication traffic. The parameters of four representative QKD systems are given in Table \ref{tbl-QKD system detail arguments}, which are used to compute the corresponding count rate of these systems by Eq.(\ref{eq:count rate q}). Upon the calculated count rates, the optimal code alphabet sizes are suggested for them in Table \ref{tbl-QKD system recommended k} according to their available storages. The theoretical $n$ is calculated by $n=2^k$ without considering the  constraint of storage, where $k$ is obtained by Algorithm \ref{Algo:1}, and the corresponding compression efficiency is named theoretical $f$. While the recommended $n$ is calculated after taking into account the storage constraint, and the corresponding compression efficiency is named as actual $f$. The systems of \cite{dixon2008gigahertz, zhang2012real, tanaka2012high} have sufficient storage, and all of their actual compression efficiencies are near 1. However, the system of \cite{stucki2009high} just has 32Kb storage for sifting which is not enough for the theoretical optimal $n=2^{22}$. The protocol adopted by \cite{stucki2009high} is coherent one-way (COW)\cite{stucki2005fast}, which needs two bits to describe each $\text{original\_key}_A$. One bit indicates whether it is a decoy or a signal, and the other determines its value when it is a signal. Hence Alice should store two bits for each $\text{original\_key}_A$. If the memory is extended to 8Mb, then the compression efficiency of the system would be 1.06. Otherwise, if 8Kb is allocated to basis sifting step, then the rest 24Kb is for bit sifting. Therefore the possible maximum code alphabet size $n_{max}$ is set as $12*2^{10}$, i.e. 12K. According to Algorithm \ref{Algo:2}, the optimal code alphabet size $n$ and optimal solution is $12*2^{10}$ and 70.57, respectively. It can be seen that the performance falls sharply due to the lack of the storage resource.

\section{\label{sec:Conclusion}Conclusion}
In this paper,  an efficient bit sifting scheme for QKD is proposed, whose core is a modified zero run length source coding algorithm with performance near Shannon's limit. The existence of optimal codelength of the source coding algorithm is proved, and a fast iteration algorithm is presented to solve the optimal parameter. Both the theoretical analysis and the experimental results demonstrate that our scheme can reduce the classical communication traffic greatly and hereby save the secure key consumption for authentication evidently. As a fast bit sifting scheme, the storage pressure of Alice can be relieved greatly by sifting out the undetected original keys in time. The impact of storage resource of a QKD system on the application of our scheme is also discussed. Some recommendations on how to apply our scheme into four representative QKD systems are given.

\appendix[\label{sec:Appendixes Proof of Lemma}Proof of Theorem \ref{Thm-R property}]
\begin{IEEEproof}
For convenience, we denote $p=1-q$, then $p \in \left[0.9, 1\right)$ and $q=1-p$. The $\overline{L}\left(z\right)$ in Eq.(\ref{eq:L_z}) can be rewritten as
\[\overline{L}\left( z \right) = \frac{{\left( {1 - p} \right)z}}{{1 - {p^{{2^z} - 1}}}},\]
and the partial derivative of the expected codelength $\overline{L}$ with respect to the variable $z$ is given by
\begin{equation}
\frac{{\partial \overline{L}}}{{\partial z}} = \frac{{(1 - p)p}}{{{{\left( {p - {p^{{2^z}}}} \right)}^2}}}\left( {p - {p^{{2^z}}} + z{2^z}{p^{{2^z}}}\ln 2\ln p} \right).
\label{eq:A-1}
\end{equation}
Due to $p \in [0.9,1)$,
\[\frac{{(1 - p)p}}{{{{\left( {p - {p^{{2^z}}}} \right)}^2}}} > 0.\]
we only need to focus on the sign of
\begin{equation}
r\left( z \right) = p - {p^{{2^z}}} + z{2^z}{p^{{2^z}}}\ln 2\ln p.
\label{eq:A-2}
\end{equation}

The partial derivative of $r\left( z \right)$ with respect to the variable $z$ can be evaluated as
\begin{equation}
\frac{{\partial r}}{{\partial z}} = z{2^z}{p^{{2^z}}}{\ln ^2}2\ln p\left( {1 + {2^z}\ln p} \right).
 \label{eq:A-3}
\end{equation}
Due to $z{2^z}{p^{{2^z}}}{\ln ^2}2\ln p < 0$, the sign of $\frac{{\partial r}}{{\partial z}}$ is determined by the sign of the expression $1 + {2^z}\ln p$. Let $1 + {2^z}\ln p > 0$, then
\[z <  - {\log _2}\left( { - \ln p} \right) \buildrel \Delta \over = {z_m}.\]
So
\begin{equation}
\left\{ {\begin{array}{*{20}{l}}
{\frac{{\partial r}}{{\partial z}} < 0,}&{{{when}}\;1 \le z{\rm{ }} < {\rm{  }}{z_m}}\\[1mm]
{\frac{{\partial r}}{{\partial z}} = 0,}&{{{when}}\;z{\rm{ }} = {\rm{  }}{z_m}}\\[1mm]
{\frac{{\partial r}}{{\partial z}} > 0,}&{{{when}}\;z{\rm{ }} > {\rm{  }}{z_m}}
\end{array}} \right..
\end{equation}
Therefore function  $r\left( z \right)$ is monotonic decreasing in the domain $z \in \left[1, z_m\right)$ and monotonic increasing in domain $\left(z_m, +\infty\right)$, and reaches the global minimum value at the point $z = z_m$, which is
\begin{equation}
r\left( {{z_m}} \right) = \frac{{ - 1 + ep + \ln \left( { - \ln p} \right)}}{e}.
\label{eq:A-4}
\end{equation}
Since
\[\frac{{\partial r\left( {{z_m}} \right)}}{{\partial p}} = {\rm{ }}1 + \frac{1}{{ep\ln p}} < -2.87, \rm{ }\forall p \in [0.9,1),\]
function $r\left(z_m\right)$ is monotonic decreasing with respect to the variable $p$ and comes to the maximum value at $p=0.9$, which is about $-0.30$. So
\begin{eqnarray}
r\left( {{z_m}} \right) < 0, \forall p \in [0.9,1).
\label{eq:A-5}
\end{eqnarray}

In addition, as $z$ approaches $ + \infty $, the limit of $r\left( z \right)$ is
\begin{eqnarray}
\mathop {\lim }\limits_{z \to  + \infty } r\left( z \right) = p, \forall p \in [0.9,1).
\label{eq:A-6}
\end{eqnarray}

Upon Eq.(\ref{eq:A-5}), Eq.(\ref{eq:A-6}) and the continuity of $r\left( z \right)$ in the domain $z \in \left[ {z_m, + \infty } \right)$, it can be inferred that there exists at least one ${z_0} \in \left( { z_m, + \infty } \right)$ satisfying that $r\left( {{z_0}} \right) = 0$ according to the intermediate value theorem\cite{johnsonbaugh2012foundations}. Besides, since $r\left( z \right)$ is monotonic increasing function in the range $z \in \left[ {z_m, + \infty } \right)$, the root of $r(z)=0$ is unique, and
\begin{eqnarray}
\left\{ {\begin{array}{*{20}{l}}
{r\left( z \right) < 0,}&{when\;z \in \left[ {{z_m},{z_0}} \right)}\\
{r\left( z \right) = 0,}&{when\; z = z_0}\\
{r\left( z \right) > 0,}&{when\;z \in \left( {{z_0}, + \infty } \right)}
\end{array}} \right.
\label{eq:A-r-km-infinity}
\end{eqnarray}

Since
\[{z_m} \ge {\left. { - {{\log }_2}\left( { - \ln p} \right)} \right|_{p = 0.9}} > 3.24,\]
we have to discuss the sign of $r\left( z \right)$ in the domain $z \in \left[ {1, z_m } \right)$. Now our concern is the sign of
\begin{equation}
r\left( 1 \right) = p - {p^2} + 2{p^2}\ln 2\ln p.
\label{eq:A-7}
\end{equation}
Th partial derivative of $r\left( 1 \right)$ with respect to the variable $p$ is given by
\begin{equation}
\frac{{\partial r\left( 1 \right)}}{{\partial p}}= 1 - 2p + 2p\ln 2 + 4p\ln 2\ln p,
\label{eq:A-8}
\end{equation}
and the 2nd partial derivative of $r\left( 1 \right)$ with respect to the variable $p$ is given by
\begin{equation}
\frac{{{\partial ^2}r\left( 1 \right)}}{{\partial {p^2}}} =  - 2 + 6\ln 2 + 4\ln 2\ln p.
\label{eq:A-9}
\end{equation}
Obviously in the domain $p \in [0.9,1)$, the function $\frac{{{\partial ^2}r\left( 1 \right)}}{{\partial {p^2}}}$ is monotonic increasing with respect to the variable $p$, and reaches the minimum value at $p = 0.9$, which is about 1.87. Thus the function $\frac{{\partial r\left( 1 \right)}}{{\partial p}}$ with respect to the variable $p$ is also monotonic increasing and also reaches the minimum value at $p = 0.9$, which is about 0.18. Therefore $r\left( 1 \right)$ is a monotonic increasing function with respect to the variable $p$ and reaches the supremum 0 at $p = 1$, so
\begin{equation}
r\left( 1 \right) < 0, \forall p \in [0.9,1).
\label{eq:A-10}
\end{equation}
Since $r\left( z \right)$ is a monotonic decreasing function in the range $z \in \left[1, z_m\right)$, we have
\begin{eqnarray}
r\left( z \right) \le r\left( 1 \right) < 0, \forall z \in \left[1, z_m\right).
\label{eq:A-r-1-km}
\end{eqnarray}

Combining Eq.(\ref{eq:A-r-km-infinity}) and Eq.(\ref{eq:A-r-1-km}), we have
\begin{equation}
\left\{ {\begin{array}{*{20}{l}}
{r\left( z \right) < 0,}&{when\;z \in \left[ {{1},{z_0}} \right)}\\
{r\left( z \right) = 0,}&{when\; z = z_0}\\
{r\left( z \right) > 0,}&{when\;z \in \left( {{z_0}, + \infty } \right)}
\end{array}} \right..
\end{equation}
Since the sign of $\frac{{\partial \overline{L}}}{{\partial z}}$ is the same as the sign of $r\left( z \right)$,
\[\left\{ {\begin{array}{*{20}{l}}
{\frac{{\partial \overline{L}}}{{\partial z}} < 0,}&{when\;z \in \left[ {1,{z_0}} \right)}\\[1mm]
{\frac{{\partial \overline{L}}}{{\partial z}} = 0,}&{when\;z = z_0}\\[1mm]
{\frac{{\partial \overline{L}}}{{\partial z}} > 0,}&{when\;z \in \left( {{z_0}, + \infty } \right)}
\end{array}} \right.\]
where ${z_0} \in \left( { z_m, + \infty } \right)$, i.e. \[{z_0} \in \left( { - {{\log }_2}\left( { - \ln \left(1-q\right)} \right), + \infty } \right).\]
\end{IEEEproof}

\section*{Acknowledgment}
We thank Z. Li in the Peking university for the discussions of the count rate of QKD system and the COW protocol, and thank Q. Zhao in the Harbin Institute of Technology for the discussions of the inference of some formulas. This work is supported by the National Natural Science Foundation of China (Grant Number:
61301099, 61361166006) and the Fundamental Research Funds for the Central Universities (Grant Number:
HIT. NSRIF. 2013061, HIT. KISTP. 201416, HIT. KISTP. 201414).

\ifCLASSOPTIONcaptionsoff
  \newpage
\fi



\bibliographystyle{IEEEtran}
\bibliography{IEEEabrv,IEEEtran}

%








\end{document}